\def\be{\begin{equation}}
\def\ee{\end{equation}}
\def\ba{\begin{array}}
\def\ea{\end{array}}

\def\Rb{{I\!\! R}}

\def\Rb{{I\!\! R}}
\def\qed{\leavevmode\unskip\penalty9999 \hbox{}\nobreak\hfill
     \quad\hbox{\leavevmode  \hbox to.77778em{%
               \hfil\vrule   \vbox to.675em%
               {\hrule width.6em\vfil\hrule}\vrule\hfil}}
     \par\vskip3pt}

\documentclass[12pt]{article}
\usepackage[lmargin=25mm,top=25mm,bottom=20mm,left=20mm,right=20mm]{geometry}
\usepackage{amsmath}
\usepackage{indentfirst}
\usepackage{CJK}
\usepackage{graphicx}

\title{Time optimal quantum control of two-qubit systems}

\begin{document}
\date{}
\setlength{\baselineskip}{22pt}
\parskip=4pt
\parindent=18pt
\maketitle
\vspace{-1cm}

\centerline{{LI Bin$^{1,2}$}, {YU ZuHuan$^{1}$}, FEI ShaoMing$^{1,3}$ and { LI-JOST XianQing$^{3}$}}

\centerline{$^1$School of Mathematical Sciences, Capital Normal
University, Beijing 100048, China;}

\centerline{$^2$School of Mathematics and Statistics, Northeast
Normal University, Changchun 130024, China;}

\centerline{$^3$Max-Planck-Institute for Mathematics in the
Sciences, Leipzig 04103, Germany}

\bigskip
\medskip

\begin{center}
\begin{minipage}{6.2in}~~~~
We study the optimal quantum control of heteronuclear two-qubit systems
described by a Hamiltonian containing both nonlocal internal drift and local control terms.
We derive an explicit formula to compute the minimum time required
to steer the system from an initial state to a specified final state.
As applications the minimal time to implement Controlled-NOT gate, SWAP gate and Controlled-U gate
is calculated in detail. The experimental realizations of these quantum gates are explicitly presented.
\end{minipage}
\end{center}
\bigskip
\smallskip

{Key wards: Time optimal quantum control, heteronuclear two-spin system, local invariants}
\medskip

\bigskip
\medskip

\section{Introduction}

The optimal control of quantum systems \cite{1,2,3} plays important roles in quantum computation
and quantum information processing \cite{Nielsen}.
For instance, the nuclear magnetic resonance (NMR) used in information processing
relies on a limited set of control variables in order to create desired
unitary transformations that manipulate
an ensemble of nuclear spins to transfer coherence between coupled spins in multidimensional
NMR-experiments \cite{5}, or to implement quantum-logic gates in NMR quantum computers \cite{8}.
There have been many rigorous results on the optimal control of spin systems from numerical calculations,
together with some experimental realizations in NMR systems \cite{a1,a2,a3,a4,a5}.

Nevertheless, it has been still a challenging problem to determine
the minimum time analytically for the implementation of an arbitrary given unitary transformation.
Based on Cartan decomposition of unitary operators,
the authors in ref. \cite{NRS} studied the minimum time required
to steer the system from some initial sate to a specified final state
for a given controllable right invariant system, described by a Hamiltonian containing both a
nonlocal internal or drift term, and a local control term.
An elegant analytical characterization of such time optimal
control in spin systems has been presented.
However, since the Cartan decomposition of a unitary operator is not unique,
the formula given in ref. \cite{NRS} can not be operationally applied to
compute the minimal time for a detailed given unitary operator.

In this paper, by using the local invariants associated with
the local equivalent transformation of unitary operators \cite{Y,JJSK},
we give an operational approach to compute the minimal time
required to implement a given unitary operator for the heteronuclear system \cite{6}.
For examples, we explicitly compute the minimal time for several
important quantum gates such as controlled-NOT, SWAP and
controlled-U ones. Moreover, based on the optimal Cartan decomposition
of these unitary operators in the derivation of the minimal time,
we get the corresponding ways to
to realize these quantum gates experimentally, with the
control Hamiltonian explicitly given.

The state of a quantum system is described by a density matrix $\rho$.
The state $\rho(0)$ at time zero evolves into the state $\rho(t)$ at time $t$,
$\rho(t)=U(t)\rho(0)U^\dag(t)$ for some unitary operator
$U(t)$. The unitary operator $U(t)$ is determined by the Hamiltonian
of the system $H(t)$ satisfying the time-dependent Schr\"odinger equation,
\be\label{ut}
\dot{U}(t)=-iH(t)U(t),
\ee
with $U(0)=I$ the identity operator.
For finite-dimensional quantum systems, $H(t)$ is a Hermitian matrix of the form,
\be\label{ht}
H(t)=H_d+\sum_{i=1}^m v_i(t)H_i,
\ee
where $H_d$ is called the drift Hamiltonian which is internal to the system, and
$\sum_{i=1}^m v_i(t)\,H_i$ is the control Hamiltonian such that the coefficients
$v_i(t)$ can be externally manipulated \cite{NRS}.

The key problem in optimal time control of a quantum system is to
find the minimal time $t^\ast$ required for the system to reach the final
state $\rho(t^\ast)$ from a initial state $\rho(0)$, namely, to implement a
unitary operator $U(t^\ast)$.

The problem can be investigated according to the algebraic
properties related to the unitary group actions. Let $G$ be a Lie
group and ${g}$ its corresponding Lie algebra. Let $K$ denote a
compact closed subgroup of $G$, and ${l}$ the Lie algebra of right
invariant vector fields on $K$. There is an one-to-one correspondence
between the vector fields $T_e(G)$ and the tangent spaces
$T_e(K)$, denoted by ${g}$ and ${l}$ respectively,
${g}={l}\oplus{p}$, ${p}={l}^{\perp}$. For a real semi-simple Lie
algebra ${g}$, one has a Cartan decomposition, $[{l},{l}]\subset
{l}$, $[{p}, {l}]={p}$, $[{p},{p}]\subset {l}$. If ${s}$ is a
subalgebra of ${g}$ contained in ${p}$, then ${s}$ is Abelian as
$[{p},{p}]\subset {l}$. A maximal Abelian subalgebra contained in
${p}$ is called a Cartan subalgebra. The homogeneous coset space
$G/K$ is a differential manifold. The Lie group $G$ has similarly a
Cartan decomposition, $G=K\,e^{{s}}\,K$.

\section{Heteronuclear two-spin system}

We consider the typical and most interesting optimal time control
problem of a heteronuclear two-spin (two-qubit) system \cite{6},
with the Hamiltonian (\ref{ht}) given by \be\label{ht2q} \ba{l}
H_d=\displaystyle\frac{\pi}{2}J\sigma_z^1\sigma_z^2,\\[3mm]
H_1=\pi\sigma_x^1,~~H_2=\pi\sigma_y^1,~~H_3=\pi\sigma_x^2,~~H_4=\pi\sigma_y^2,
\ea
\ee
where $\sigma_x^\alpha$, $\sigma_y^\alpha$ and $\sigma_z^\alpha$ are Pauli matrices acting on
the $\alpha$th quibt, $\alpha=1,2$ and $J$ is the coupling constant of the system.

In this case the problem is related to the special unitary group
$G=U(4)$. As an arbitrary two-qubit gate can be decomposed as the
product of a gate $U_1\in SU(4)$ and a global phase shift
$e^{i\theta}$, $\theta\in \Rb$, the problem is reduced to the study of the
group $SU(4)$ in stead of the group $U(4)$. The Lie algebra $su(4)$ of $SU(4)$ has a
Cartan decomposition ${g}={p}\oplus{l}$ with $
{l}=span\frac{i}{2}\{\sigma_x^1,\sigma_y^1,\sigma_z^1,\sigma_x^2,\sigma_y^2,\sigma_z^2\}
$ and
$$
{p}=span\frac{i}{2}\{\sigma_x^1\sigma_x^2,\sigma_x^1\sigma_y^2,\sigma_x^1\sigma_z^2,\sigma_y^1\sigma_x^2,
\sigma_y^1\sigma_y^2,\sigma_y^1\sigma_z^2,\sigma_z^1\sigma_x^2,\sigma_z^1\sigma_y^2,\sigma_z^1\sigma_z^2\},
$$
together with the Cantan subalgebra, $
{s}=span\frac{i}{2}\{\sigma_x^1\sigma_x^2,\sigma_y^1\sigma_y^2,\sigma_z^1\sigma_z^2\}.
$

Since the set of all the local gates $K$ is a connected Lie group
$SU(2)\otimes SU(2)$ in $SU(4)$,
${l}=span\frac{i}{2}\{\sigma_x^1,~\sigma_y^1,~\sigma_z^1,~\sigma_x^2,~\sigma_y^2,~\sigma_z^2\}$
is just the Lie subalgebra corresponding to $K$. Therefore $U\in
SU(4)$ can be decomposed as: \be\label{dec}
U=k_1\exp\{\frac{i}{2}(c_1\sigma_x^1\sigma_x^2+c_2\sigma_y^1\sigma_y^2+c_3\sigma_z^1\sigma_z^2)\}k_2,
\ee where $k_1,~k_2\in SU(2)\otimes SU(2),$ and $c_1,~c_2,~c_3\in
\Rb.$

When the control terms in the Hamiltonian are large enough
compared with the internal couplings, any single-qubit operation can be achieved
almost instantaneously. It has been proved in \cite{NRS} that for
the Hamiltonian system described by eq. (\ref{ht2q}), the minimal time to implement
a quantum gate $U$ of the form (\ref{dec}) is given by
$$
t^\ast=\frac{1}{\pi J}\min \,\sum_{i=1}^3 c_i,~~~~~c_i>0.
$$
Since for given $U$, its decompositions of the form (\ref{dec}) are not unique,
it is a challenging problem to find the minimum of $\sum_{i=1}^3 c_i$.

To find an analytical formula of $t^\ast$, we consider the
local invariants and local equivalent classes in $U(4)$.
Two unitary transformations $U,~U_1\in U(4)$ are said to be
locally equivalent if they satisfy,
$U=k_1\,U_1\,k_2$, for some $k_1,~k_2\in U(2)\otimes U(2)$, which
defines a set of invariants under such equivalent transformations.
These invariants can be expressed as \cite{JJSK},
\begin{equation}\label{g1g2}
G_1=\frac{tr^2[m(U)]}{16\,detU},
~~~~~G_2=\frac{tr^2[m(U)]-tr[m^2(U)]}{4\,detU},
\end{equation}
where $m(U)=U_B^TU_B$, $U_B=O^\dag UO$, and
\[ \begin{split}O=\frac{1}{\sqrt{2}}\begin{pmatrix}
1 & 0 & 0 & i \\
0 & i & 1 & 0 \\
0 & i & -1 & 0 \\
1 & 0 & 0 & -i
\end{pmatrix}.
\end{split}\]

As $U$ can be expressed in the form (\ref{dec}),
one has the invariants \cite{JJSK},
\be\label{g12}
G_1=a+ib,~~~~G_2=c,
\ee
where
\begin{align}
a=&\cos^2 c_1\cos^2 c_2\cos^2 c_3-\sin^2 c_1\sin^2 c_2\sin^2 c_3, \\
b=& \frac{1}{4}\sin 2c_1\sin 2c_2\sin 2c_3,\\
c=&4\cos^2 c_1\cos^2 c_2\cos^2 c_3-4\sin^2 c_1\sin^2
c_2\sin^2c_3-\cos 2c_1\cos 2c_2\cos 2c_3.
\end{align}
Our main idea is to find the solution $c_1$, $c_2$ and $c_3$ from $G_1=a+ib$, $G_2=b$,
according to the local invariants $a$, $b$ and $c$,
so that the value $\sum_{i=1}^3 c_i$ will be independent of the detailed Cartan expression (\ref{dec}).

It is direct to verify that
$\sqrt{a^2+b^2}=\cos^2 c_1\cos^2 c_2\cos^2 c_3+\sin^2 c_1\sin^2
c_2\sin^2 c_3$. Therefore we have
\begin{eqnarray}{l}
\cos^2 c_1\cos^2 c_2\cos^2 c_3=\frac{1}{2}(\sqrt{a^2+b^2}+a),\label{11a}\\
\sin^2 c_1\sin^2 c_2\sin^2 c_3=\frac{1}{2}(\sqrt{a^2+b^2}-a).\label{11b}
\end{eqnarray}
However, by using the formula $\cos^2\alpha+\sin^2\alpha=1$, from (\ref{11a})
we also have
$$
\ba{rcl}
\cos^2 c_1\cos^2 c_2\cos^2 c_3
&=&1-(\sin^2 c_1+\sin^2 c_2+\sin^2 c_3)-\sin^2c_1\sin^2 c_2\sin^2c_3\\[2mm]
&&+(\sin^2 c_1\sin^2 c_2+\sin^2c_1\sin^2 c_3+\sin^2 c_2\sin^2
c_3).
\ea
$$
Hence one gets
\begin{equation}\label{13}
(\sin^2 c_1+\sin^2 c_2+\sin^2 c_3)-(\sin^2
c_1\sin^2 c_2+\sin^2c_1\sin^2 c_3+\sin^2 c_2\sin^2
c_3)=1-\sqrt{a^2+b^2}.
\end{equation}

Moreover, eq. (\ref{11b}) can be written as
$\cos 2c_1\cos 2c_2\cos 2c_3=4a-c$. While
$$
\ba{rcl}
\cos2 c_1\cos2 c_2\cos2 c_3&=&(1-2\sin^2 c_1)(1-2\sin^2c_2)(1-2\sin^2 c_3) \\[2mm]
&=&1-2(\sin^2 c_1+\sin^2 c_2+\sin^2 c_3)-8\sin^2c_1\sin^2
c_2\sin^2c_3\\[2mm]
&&+4(\sin^2 c_1\sin^2 c_2+\sin^2c_1\sin^2 c_3+\sin^2 c_2\sin^2
c_3).
\ea
$$
Therefore we obtain
\begin{equation}\label{14}
(\sin^2 c_1+\sin^2 c_2+\sin^2 c_3)-2(\sin^2
c_1\sin^2 c_2+\sin^2c_1\sin^2 c_3+\sin^2 c_2\sin^2
c_3)=\frac{1+c}{2}-2\sqrt{a^2+b^2}.
\end{equation}
From eqs. (\ref{13}) and (\ref{14}), we have
\begin{equation}\label{15}
\left\{\begin{aligned}
&\sin^2 c_1+\sin^2 c_2+\sin^2 c_3 =1+\frac{1-c}{2},\\
&\sin^2 c_1\sin^2 c_2+\sin^2c_1\sin^2 c_3+\sin^2
c_2\sin^2c_3
&=\sqrt{a^2+b^2}+\frac{1-c}{2}.
                          \end{aligned} \right.
                          \end{equation}

From eqs. (\ref{15}) and (\ref{11b}), we see that $\sin^2 c_1$, $\sin^2
c_2$ and $\sin^2 c_3$ can be considered as the solutions of the following cubic equation,
$$
x^3+px^2+qx+r=(x-\sin^2 c_1)(x-\sin^2 c_2)(x-\sin^2 c_3)=0,
$$
where
\begin{equation}
\ba{l}\displaystyle
p =-(1+\frac{1-c}{2}),~~~
q=\sqrt{a^2+b^2}+\frac{1-c}{2},~~~
\displaystyle
r=-\frac{1}{2}(\sqrt{a^2+b^2}-a).
\ea
\end{equation}
Set $X=x+p/3$. The cubic equation becomes
\be\label{18}
X^3+PX+Q=0,
\ee
where
\begin{equation}\label{pq}
\left\{ \begin{aligned}
&P=q-\frac{p^2}{3}=-\frac{1}{12}(c^2+3)+\sqrt{a^2+b^2},\\
&Q=\frac{2p^3}{27}-\frac{pq}{3}+r=\frac{1}{108}(c^3-9c)+\frac{a}{2}-\frac{c}{6}\sqrt{a^2+b^2}.
\end{aligned} \right.
\end{equation}

To deal with eq. (\ref{18}), we consider two elementary functions
$Y=X^3$ and $Y=-PX-Q$ in $\Rb^2$. These two curves may intersect at
one, two or three points with respect to different values of $P$ and
$Q$. First, for the most special case: half the curve $Y=X^3$
tangents to the line $Y=-PX-Q$, one has one single real solution
$X_1$ and another two real ones $X_2=X_3$,
$$
\ba{rcl}
X^3+PX+Q&=&(X-X_1)(X-X_2)^2\\[2mm]
&=&X^3-(X_1+2X_2)X^2+(2X_1X_2+X_2^2)X-X_1X_2^2=0,
\ea
$$
and also $X_1+2X_2 =0$, $2X_1X_2+X_2^2=P$, $-X_1X_2^2=Q$.
When $P$ and $Q$ satisfy the condition $P^3/27+Q^2/4=0$, the
solutions of eq. (\ref{18}) are
$$
X_1=-2(Q/2)^{1/3},~~~~X_2=X_3=(Q/2)^{1/3}.
$$

Second, eq. (\ref{18}) has three different real solutions when
the inequality $P^3/27+Q^2/4<0$ is satisfied,
$$
X_1=-\frac{2\sqrt{-3P}}{3}\cos\frac{\theta}{3};~~~
X_2,X_3=-\frac{2\sqrt{-3P}}{3}\cos\frac{\theta}{3}\pm\frac{\sqrt{3}}{3}\sin\frac{\theta}{3},
$$
where $\theta=arccos T$ and $T=27Q/(2(-3P)^{3/2})\in(-1,~1)$.

According to the Shengjin's formulas, eq. (\ref{18}) may have one
single real solution and two imaginary solutions when the inequality
$P^3/27+Q^2/4>0$ holds. Since in our case
$P^3/27+Q^2/4\leq0$ is always satisfied, there will be no
imaginary solutions.

Combining the above results, we have

{\bf [Theorem]}~~For the system (\ref{ht}) with the two-qubit Hamiltonian given by eq. (\ref{ht2q}),
the minimal time to implement a unitary operator $U$ is given by
\be\label{thm}
t^\ast(U)=\frac{1}{\pi J}\min\, \sum_{i=1}^3 c_i=\frac{1}{\pi J}\sum_{i=1}^3\arcsin\sqrt{X_i+\frac{3-c}{6}},
\ee
where
$$X_1=-2(\frac{Q}{2})^{\frac{1}{3}},~~~X_2=X_3=(\frac{Q}{2})^{\frac{1}{3}},$$
if ${P^3}/{27}+{Q^2}/{4}=0$; and
$$
X_1=-\frac{2\sqrt{-3P}}{3}\cos\frac{\theta}{3};~~~
X_2,X_3=-\frac{2\sqrt{-3P}}{3}\cos\frac{\theta}{3}\pm\frac{\sqrt{3}}{3}\sin\frac{\theta}{3},
$$
if ${P^3}/{27}+{Q^2}/{4}<0$.

\section{Applications}

We have presented an analytical formula to compute the minimal time
required to implement an arbitrary given unitary operation for
two-qubit system (\ref{ht2q}). Two-qubit operations are the most
fundamental ones in quantum computation and quantum information
processing. As examples, here we compute the minimal time for
several important two-qubit gates.

{\it Example 1.} \emph{Controlled-NOT gate}~~~ $C_{NOT}=|0\rangle\langle0|\otimes I_2
+|1\rangle\langle1|\otimes (|1\rangle\langle0|+|0\rangle\langle1|)$, where $I_2$ is the $2\times 2$ identity matrix.
From eq. (\ref{g1g2}) we obtain $G_1=0$, $G_2=1$. That is,
$a=b=0$, $c=1$ due to eq. (\ref{g12}). Hence from eq. (\ref{pq}) we have $P=-{1}/{3}$, $Q=-{2}/{27}$.
We have ${P^3}/{27}+{Q^2}/{4}=0$,
$X_1={2}/{3}$, $X_2=X_3=-{1}/{3}$, and
$c_1={\pi}/{2}$, $c_2=c_3=0$. Therefore
the minimal time is given by $t^*(C_{NOT})=\frac{1}{\pi J}\sum_{i=1}^3c_i=\frac{1}{2J}$.

To optimally implement the gate $C_{not}$ experimentally, one has to find the
Cartan decomposition of $C_{not}$ which fulfils $t^*(C_{NOT})=\frac{1}{2J}$.
Let us assume
$$
\exp(\frac{\pi i}{4})Cnot=k_1exp(\frac{\pi
i}{4}\sigma_x\otimes\sigma_x)k_2
$$
for some $k_1,k_2\in SU(2)\otimes SU(2)$. Note that
$exp(\frac{\pi i}{4}\sigma_x\otimes\sigma_x)=\frac{\sqrt
2}{2}(I_2+i\sigma_x\otimes \sigma_x)$. The problem is to
compute $k_1$ and $k_2$ in the following equation,
\be\label{pp}
(1+i)\left(
         \begin{array}{cc}
           I_2 & 0 \\
          0 & \sigma_x \\
         \end{array}
       \right)=k_1\left(
                    \begin{array}{cc}
                      I_2 & i\sigma_x \\
                      i\sigma_x & I_2 \\
                    \end{array}
                  \right)k_2.
\ee
Set $k_1=A\otimes B$, $k_2=C\otimes D$, with $A=(a_{ij})$,
$C=(c_{ij})$, $B$, $D\in SU(2)$.
A direct computation yields
$$
k_1\left(\begin{array}{cc}
                      I_2 & i\sigma_x \\
                      i\sigma_x & I_2 \\
                    \end{array}
                  \right)k_2=\left(
                               \begin{array}{cc}
                                 f_{11} & f_{12} \\
                                 f_{21} & f_{22} \\
                               \end{array}
                             \right),
$$
where
$$
\ba{l}
f_{11}=(a_{11}c_{11}+a_{12}c_{21})\,BD+i(a_{12}c_{11}+a_{11}c_{21})\,B\sigma_xD,\\[2mm]
f_{12}=(a_{11}c_{12}+a_{12}c_{22})\,BD+i(a_{12}c_{12}+a_{11}c_{22})\,B\sigma_xD,\\[2mm]
f_{21}=(a_{21}c_{11}+a_{22}c_{21})\,BD+i(a_{22}c_{11}+a_{21}c_{21})\,B\sigma_xD,\\[2mm]
f_{22}=(a_{21}c_{12}+a_{22}c_{22})\,BD+i(a_{22}c_{12}+a_{21}c_{22})\,B\sigma_xD.
\ea
$$
From eq. (\ref{pp}) one has $f_{12}=f_{21}=0$, namely
$$
\ba{l}
(a_{11}c_{12}+a_{12}c_{22})\,I_2+i(a_{12}c_{12}+a_{11}c_{22})\,\sigma_x=0,\\[2mm]
(a_{21}c_{11}+a_{22}c_{21})\,I_2+i(a_{22}c_{11}+a_{21}c_{21})\,\sigma_x=0.
\ea
$$
By detailed analysis one obtains
$$C=\frac{\sqrt 2}{2}\left(
      \begin{array}{cc}
        \exp(i\theta) & -\exp(-i\theta)  \\
        \exp(i\theta)  & \exp(-i\theta)  \\
      \end{array}
    \right)~~~
\text{and}~~
A=\frac{\sqrt 2}{2}\left(
      \begin{array}{cc}
        \exp(i\beta) & \exp(i\beta)  \\
        -\exp(-i\beta)  & \exp(-i\beta)  \\
      \end{array}
    \right).
$$
From the expressions of $f_{11}$ and $f_{22}$ we have further
$$
D=\left(
       \begin{array}{cc}
         0 & 1 \\
         -1 & 0 \\
       \end{array}
     \right)~~~
\text{and}~~
B=\frac{\sqrt{2}}{2}(I_2+i\sigma_x)\left(
       \begin{array}{cc}
         0 & 1 \\
         -1 & 0 \\
       \end{array}
     \right).
$$
Therefore the Cartan decomposition of $C_{not}$ reads,
\be\label{cd}
\ba{rcl}
Cnot&=&\displaystyle\exp(-\frac{\pi i}{4})\left(\exp(-\frac{\pi i}{4}\sigma_y)\exp(\frac{\pi i}{4}\sigma_x)\otimes \exp(\frac{\pi i}{4}\sigma_x)
\exp(\frac{\pi i}{2}\sigma_y)\right)\cdot\\[3mm]
&&\displaystyle\exp(\frac{\pi i}{4}\sigma_x\otimes \sigma_x)\left(\exp(\frac{\pi i}{4}\sigma_y)\otimes \exp(-\frac{\pi i}{2}\sigma_y)\right).
\ea
\ee

To find the detailed way to implement $C_{not}$ experimentally, we
expand the factor $\exp(\frac{\pi i}{4}\sigma_x\otimes \sigma_x)$ by using the following formula,
\be\label{p1}
\exp(\frac{i}{4}\sigma_x\otimes\sigma_x)=\left(\exp(-\frac{\pi
i}{4}\sigma_y)\otimes\exp(\frac{\pi i}{4}\sigma_y)\right)\exp(-\frac{\pi i}{4}\sigma_z\otimes\sigma_z)\left(\exp(\frac{\pi
i}{4}\sigma_y)\otimes\exp(\frac{-\pi i}{4}\sigma_y)\right).
\ee
Denote $T_m=\sigma_m\otimes 1$ and $S_m=1\otimes\sigma_m, m=x,y,z$.
We can rewrite the Cartan decomposition of $C_{not}$ as,
$$
C_{not}=\exp(\frac{-\pi}{4}i)\exp(-\frac{\pi}{4}iT_y)\exp(\frac{\pi}{4}i(T_x+S_x))\exp(-\frac{\pi}{4}i(T_y+S_y))\exp(-\frac{\pi}{4}iT_zS_z)
\exp(\frac{\pi}{4}i(2T_y+S_y)).
$$
From the Sch\"{o}dinger equation (\ref{ut}) and the Hamiltonian (\ref{ht}), (\ref{ht2q}),
we see that the unitary operator $C_{not}$ can be implemented, up to a global phase, by manipulating the control
Hamiltonian such that
$$
H(t)=\left\{
          \begin{array}{ll}\displaystyle
            H_d-\frac{N}{2}(H_2+\frac{H_4}{2}), &~~~\displaystyle {t\in [0,\frac{1}{N}];} \\[3mm]\displaystyle
            H_d, &~~~\displaystyle {t\in[\frac{1}{N},\frac{1}{N}+\frac{1}{2J}];} \\[3mm]\displaystyle
            H_d+\frac{N}{4}(H_2+H_4), &~~~\displaystyle {t\in[\frac{1}{N}+\frac{1}{2J},\frac{2}{N}+\frac{1}{2J}];}  \\[3mm]\displaystyle
           H_d-\frac{N}{4}(H_1+H_3), &~~~\displaystyle {t\in[\frac{2}{N}+\frac{1}{2J},\frac{3}{N}+\frac{1}{2J}];}  \\[3mm]\displaystyle
            H_d+\frac{N}{4}H_2, &~~~\displaystyle {t \in [\frac{3}{N}+\frac{1}{2J},\frac{4}{N}+\frac{1}{2J}],}
          \end{array}
        \right.
$$
where $N$ is a real parameter.

The parameter $N$ in the control Hamiltonian should be large enough,
$N\rightarrow \infty$, so that the drift Hamiltonian $H_d$ can be ignored during all the local
unitary evolutions, and the time needed for local unitary evolutions can be put to zero. The
finite time needed to implement $C_{not}$ is in the second step at time interval $t\in[\frac{1}{N},\frac{1}{N}+\frac{1}{2J}]$.
For $N\rightarrow \infty$, one reaches the optimal time $\frac{1}{2J}$.

{\it Example 2.} \emph{SWAP gate}~~~For the gate $Swap$,
$$
Swap=\frac{1}{2}\left(
              \begin{array}{cc}
                I_2+\sigma_z & \sigma_x+i\sigma_y \\
               \sigma_x-i\sigma_y & I_2-\sigma_z \\
              \end{array}
            \right),
$$
we have $G_1=-1$ and $G_2=-3$, from which we get
$a=-1$, $b=0$ and $c=-3$. According to the theorem, we obtain
$P=0$, $Q=0$. Since ${P^3}/{27}+{Q^2}/{4}=0$ in this case, we have
three real solutions $X_1=X_2=X_3=0$. Hence
$c_1=c_2=c_3={\pi}/{2}$, and $t^*(SWAP)=\frac{1}{\pi J}\sum_{i=1}^3 c_i=\frac{3}{2J}$.

Therefore the Cartan decomposition of $Swap$ is simply of the form
$$
Swap=\exp(\frac{-\pi i}{4})\exp( \frac{\pi i}{4}(\sigma_x\otimes\sigma_x+\sigma_y\otimes\sigma_y+\sigma_z\otimes\sigma_z)).
$$
From eq. (\ref{p1}) and the following formula
$$
\exp(\frac{i}{4}\sigma_y\otimes\sigma_y)=\left(
\exp(-\frac{\pi i}{4}\sigma_x)\otimes\exp(\frac{\pi
i}{4}\sigma_x)\right)
\exp(-\frac{\pi}{4}i\sigma_z\otimes\sigma_z)
\left(\exp(\frac{\pi i}{4}\sigma_x)\otimes\exp(-\frac{\pi
i}{4}\sigma_x)\right),
$$
we have
$$
\ba{rcl}\displaystyle
\exp(\frac{\pi i}{4})Swap&=&\displaystyle\exp(-i\frac{1}{4}(H_2-H_4))\exp(-i\frac{1}{2J}H_d)\cdot\\[2mm]
&&\displaystyle\exp(i\frac{1}{4}(H_2-H_4))\exp(-i\frac{1}{4}(H_1-H_3))\cdot\\[2mm]
&&\displaystyle\exp(-i\frac{1}{2J}H_d)\exp(-i(\frac{1}{4}H_3))
\exp(-i\frac{1}{2J}H_d)\exp(i\frac{1}{4}H_1).
\ea
$$
From the above expression, one can easily get the corresponding steps to implement
Swap gate by choosing the control parameters in the control Hamiltonian.

{\it Example 3.} \emph{$\sqrt{SWAP}$ gate}~~~
For this gate we have $G_1=i/4$, $G_2=0$, which yields
$a=0$, $b=1/4$, $c=0$ and $P=Q=0$. Similar to the $Swap$ gate case, one has
${P^3}/{27}+{Q^2}/{4}=0$. Hence we get the solution,
$c_1=c_2=c_3={\pi}/{4}$, and
$t^*(\sqrt{SWAP})=\frac{3}{4J}$. The gate can be implemented
according to the following decomposition,
$$
\ba{rcl}\displaystyle
\exp(\frac{\pi i}{8})\sqrt{SWAP}&=&\displaystyle\exp(-i\frac{1}{4}(H_2-H_4))\exp(-i\frac{1}{4J}H_d)\cdot\\[2mm]
&&\displaystyle\exp(i\frac{1}{4}(H_2-H_4))\exp(-i\frac{1}{4}(H_1-H_3))\cdot\\[2mm]
&&\displaystyle\exp(-i\frac{1}{4J}H_d)\exp(-i(\frac{1}{4}H_3))
\exp(-i\frac{1}{4J}H_d)\exp(i\frac{1}{4}H_1).
\ea
$$

{\it Example 4.} \emph{Controlled-U gate}~~~ The controlled-U gate
is of the form $C_U=|0\rangle\langle0|\otimes I
+|1\rangle\langle1|\otimes U$, where $U$ is an arbitrary
single-qubit unitary operation, $U=exp(i
\gamma_1\sigma_x+i\gamma_2\sigma_y+i\gamma_3i\sigma_z)$,
$\gamma_1,\gamma_3,\gamma_3\in \Rb$. The corresponding local
invariants are $G_1=\cos^2\gamma$, $G_2=2\cos^2\gamma+1$, where
$\gamma=\sqrt{\gamma_1^2+\gamma_2^2+\gamma_3^2}$. Accordingly we
have $a=\cos^2\gamma$, $b=0$ and $c=2\cos^2\gamma+1$. As in this
case one has $P=-\sin^4\gamma/{3}$ and $Q=-{2}\sin^6\gamma/{27}$,
the condition ${P^3}/{27}+{Q^2}/{4}=0$ is satisfied. Hence $X_1={2}\sin^2\gamma/{3}$ and
$X_2=X_3=-\sin^2\gamma{3}$. Therefore
$c_1=\arcsin\sqrt{X_1+({3-c})/{6}}=\arcsin|\sin\gamma|$,
$c_2=c_3=0$, and the minimal time to implement $C_U$ is
$t^*(C_U)=\frac{1}{\pi J}\arcsin|\sin\gamma|$.

\section{Discussions}

By using the local invariants of unitary operators, we have
presented an explicit formula of the minimal time required to
implement a given unitary operator for the heteronuclear two-qubit
quantum system. The formula can be easily used to compute the
minimal time needed to implement the quantum gates such as
$C_{NOT}$, SWAP and controlled-U ones. The protocols we presented
for optimally implementing the quantum gates can be directly
operated in the heteronuclear system \cite{6}. Our idea, employing
both the Cartan decomposition of a unitary operator and its local
invariants, can be also used for computing the optimal control time
for other quantum systems.

\bigskip
\noindent{\bf Acknowledgments}\, \,
This work is supported by the National Natural Science Foundation of China (Grant No. 11275131)
and the National Research Foundation for the Doctoral Program of Higher Education of China.

\end{document}